\documentclass{article}
\usepackage{waspaa21,amsmath,graphicx,url,times}
\usepackage{color}

\title{Audio Transformers}

\name{Prateek Verma and 
      Jonathan Berger}
\address{Stanford University\\ 450 Jane Stanford Way, Stanford CA, 94305, \\{prateekv, brg}@stanford.edu              
}
\usepackage{cuted}
\usepackage{capt-of}
\usepackage[symbol]{footmisc}

\begin{document}

\ninept

\maketitle
\begin{sloppy}
\begin{strip}\centering
\includegraphics[width=14cm,height=5cm]{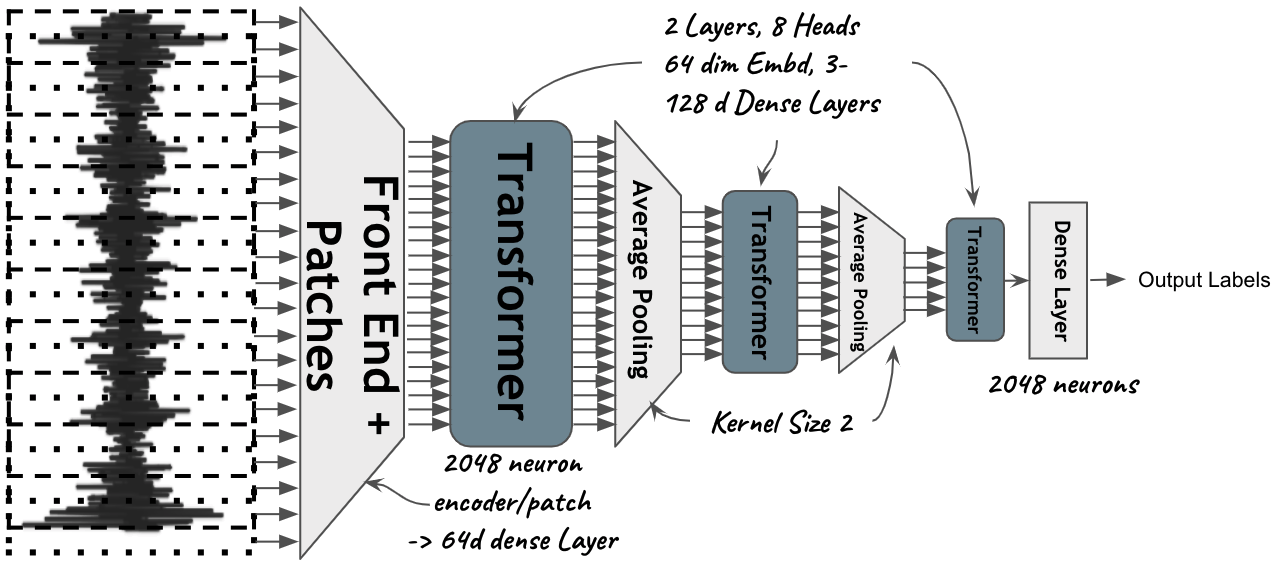}
\captionof{figure}{An overview of the proposed Audio Transformer architecture using front end fully connected encoder with Transformer layers and pooling layers. It takes 1s of input, and divides it into patches of size 25ms, followed by learning a front end, to feed it to Transformer.
\label{fig:feature-graphic}}
\end{strip}

\begin{abstract}
Over the past two decades, CNN architectures have produced compelling models of sound perception and cognition, learning hierarchical organizations of features. Analogous to successes in computer vision, audio feature classification can be optimized for a particular task of interest, over a wide variety of datasets and labels. In fact similar architectures designed for image understanding have proven effective for acoustic scene analysis. Here we propose applying Transformer based architectures without convolutional layers to raw audio signals. On a standard dataset of Free Sound 50K, comprising of 200 categories, our model outperforms convolutional models to produce state of the art results. This is significant as unlike in natural language processing and computer vision, we do not perform unsupervised pre-training for outperforming convolutional architectures. On the same training set, with respect mean average precision benchmarks, we show a significant improvement. We further improve the performance of Transformer architectures by using techniques such as pooling inspired from convolutional network designed in the past few years. In addition, we also show how multi-rate signal processing ideas inspired from wavelets, can be applied to  the Transformer embeddings to improve the results. We also show how our models learns a non-linear, non-constant bandwidth filter-bank, which shows an adaptable time frequency front end representation for the task of  audio understanding, different from other tasks e.g. pitch estimation.
\footnote{ *Although not the first instance of an acoustic scene understanding model without convolutions, this is, to our knowledge, the first end-to-end one. At the same time as viT, \cite{verma2020framework} showed how in a two-step process, one can achieve the same.}
\end{abstract}


\section{Introduction and Related Work}
\label{sec:intro}

Acoustic scene analysis is a classical signal processing and machine learning problem whose goal is to predict the contents of an input signal within a brief duration, typically one second. In addition to modeling perception, computer simulation of hearing combined with models of other sensory systems will help bridge the gap between humans and computers. For the past decade, CNNs have become a de-facto architecture in learning mappings from fixed dimensional inputs to fixed dimensional outputs \cite{he2016deep,gemmeke2017audio}. CNN architectures inspired from vision \cite{he2016deep}, adapted for acoustic scene understanding, achieve similar performance gains for audio also. 

The core backbone of this work is Transformer architecture which recently have recently produced state of the art results in a variety of domains, including protein sequences \cite{madani2020progen}, text \cite{brown2020language,devlin2018bert},  symbolic music \cite{huang2018music}, video\cite{sun2019videobert,girdhar2019video} and image understanding \cite{dosovitskiy2020image,parmar2018image}. By learning transformers on the latent representations, and conditioning a wavenet generator, they were able to achieve compelling results in music generation \cite{dhariwal2020jukebox} and style transfer, which was impossible without the guidance of meta-data and convolutional architectures \cite{verma2018neural}. They have also been used in learning latent audio representations such as \cite{verma2019neuralogram,oord2017neural} for solving pseudo-tasks such as in-filling to learn time-dependent representations \cite{verma2020framework,devlin2018bert}. As opposed to learning latent representations, the reduced time-scales of Transformers can advantageously model input representations. A major drawback of convolutional architecture is the fixed filter across the entire input. Furthermore, Transformers take advantage of attention mechanism with the output at a location dependent upon the input at some other location. 

The core idea of this work is to replace traditional convolutional based architectures \cite{gemmeke2017audio}, combined convolutional and Transformer architectures \cite{baevski2020wav2vec, baevski2019vq}, and recurrent architectures \cite{haque2018conditional,haque2019audio} with a purely Transformer based architecture. Our work is distinct from the method proposed in \cite{verma2020framework} which was not an end-to-end approach and which required a two step approach (specifically, learning a dictionary of latent codes, and using the discrete latent codes as an input to transformer architectures). Similar approaches were successfully used in areas such as speech recognition \cite{baevski2020wav2vec} to mimic BERT \cite{devlin2018bert}. All these state of the art performances were possible due to the architectures' ability to model long term dependency inputs and the attention mechanism  present in them enabling focus only on the part of the input that is important \cite{vaswani2017attention}. 

The organization of the paper is as follows: Section I introduces the problem and the literature survey followed by the dataset we used to benchmark the results in section II. The next section details the methodology followed by results and discussion in Section IV. We conclude the paper in Section V followed by our thoughts of future work and references.

\section{Dataset}
\label{sec:dataset}

We train and evaluate our architectures with FSD50K \cite{fonseca2020fsd50k}, an open dataset of over 51k audio files comprising over 100 hours of manually labeled audio using 200 classes drawn from the AudioSet \cite{gemmeke2017audio} ontology. FSD50K is freely available under the creative commons license, contains many more high quality audio annotations, and twice number of training examples in the balanced set-up than AudioSet. We used the already provided training and the validation splits to tune the model and tested them on the evaluation setup provided. In total there are about 51,197 clips available ranging from 0.3-30s. We downsample all the clips to 16kHz sampling rate using \cite{virtanen2020scipy}. We follow the same setup for reporting the results as done in \cite{fonseca2020fsd50k}. All the training was carried on 1s audio chunks with the labels inherited for all the chunks in clips greater than 1s. For samples less than 1s the audio clip is repeated to fill 1s, resulting in a single training example for that clip. On an average, the duration per clip is 7.6s, with 1.22 average labels per clip, uploaded by 7225 user ids, thus encompassing a diverse range of sources, acoustic environments, microphones, and locations, to name a few. 

\section{Methodology}
\label{sec:method}

\subsection{Baseline Transformer Architectures}
This section describes the Transformer architecture as described in \cite{vaswani2017attention} that we used to train the system as shown in Figure 1. A detailed explanation is given in \cite{opennmt}, but for the sake of clarity and completeness we describe it here. As a black-box, which we would describe in more detail in this section, it takes as an input a sequence of a fixed length $T$, and produces the same length but with a chosen dimension, which we call $E$, which denotes the size of the latent space. More specifically,  it maps a sequence $\textbf{x} = {(x_1, x_2, .... x_{T})}$ to a sequence of same length $T$,  namely $\textbf{z}: ({z_1, z_2, .... z_{T}})$ , where each of the dimensions of  $({z_1, z_2, .... z_{T}})$ is the chosen hyper-parameter $E$, which in our case is 64, the size of the embedding. For the sake of brevity, we would explain only one Transformer Encoder, and for a model with layers $L$, each of the stack is super-imposed on the other one.

Each Transformer module consists of a attention block and a feed-forward block. The output of each of them is passed through a layer norm and a residual layer. So after both the attention block and the feed-forward block, if the input to a sub-block (attention$F_a$ or feed-forward$F_{ff}$ block) is a sequence $x_b$, instead of passing the output directly  to the next module/sub-block, we pass along the block layer norm and the residual output $x_{bo}$ as $ x_{bo}=LayerNorm (x_b + F_{a/ff}(x_b))$
This follows the notion that layer-norm/skip connections help in better convergence/improved performance. We now describe each of the two sub-blocks that are part of the transformer block namely, i) multi-headed causal attention ii) feed-forward architecture

\subsubsection{Multi-Headed Causal Attention} A multi-headed causal attention function can be described as a weighting function that decides how to get the output of each step. It learns a probabilistic score of how important each of the embedding, is while predicting the output.  A multi-headed attention consists of first learning a probabilistic score. It is then multiplied with each of the inputs to determine how important each of the input is for prediction of the embedding for a position $pos$ belonging to  $1,2,3....T$. We use scaled-dot product attention as the type of attention mechanism. A query, key and a value vector is learned for each of the position for each of the inputs. This is done by implicitly learning matrices, $W_Q$, $W_K$ and $W_V$ to produce a query vector $q$, key vector $k$ and value vector $v$ for each of the inputs for a single attention head.  We take the dot product of query and key vectors, the result is multiplied by a normalization factor (the inverse of square root of size of the vector as done in \cite{vaswani2017attention}), before taking a soft-max across all the inputs. Each of the value vector is multiplied by this score to get the output of the attention module. Mathematically, for a query matrix $Q$, key matrix $K$, and a value matrix $V$, it is defined as, $Attention(Q,K,V) = softmax(\frac{QK^T}{\sqrt{d_k}})$. We can also learn multiple such attention maps for $h$ attention heads, defined as, $MutliHeadAttention(Q,K,V) = Concat(h_1,h_2,...h_h)W_o $, 
where each of the attention heads $h_i$ is defined as 
$$Attention(Q_i,K_i,V_i) = softmax(\frac{Q_iK_i^T}{\sqrt{d_k}})$$
and $W_o$ is a matrix learned during training. 

\subsubsection {Feed Forward Architecture \&  Positional Information } We weigh the saliency of each of the input signal via multi-headed attention for passing at a position $pos$. Each of the input at position $pos$ is passed through a feed-forward architecture. We have the output of the feed-forward layers $x_{bo}$ for an input $x_b$, for the dimension of feed-forward layers $d_{ff}$, in case of 2-layer network is, $ FF(x_b) = max(0,x_bW_{1}+ b_{1})W_{2} + b_2. $. We apply this function identically at each of the inputs. As described in \cite{vaswani2017attention}, to each of the inputs, positional encoding are added. As the input is passed on as a list, the model does not take into account the relative position, and thus the positional encoding are needed. For any position $pos$ for the dimension $i$ of the latent space, we use sinusoidal function, i.e. to each position $pos$ and embedding dimension $i$ in $E$, we add, $$ PE_{pos,2i/2i+1} = sin/cos(pos/10000^{(2i/E)})$$ This adds positional information for each point in time, of input with dimension $E$ , before passing thorough self-attention layers. 

\subsection{Adapting Transformer Architecture for raw waveforms}
We adapt Transformer architectures using ideas from traditional signal processing. Since the Transformer has $\mathcal{O}(n^2)$  complexity w.r.t memory and computation requirements, we choose to follow the traditional route of windowing the signal. For all the experiments, as discussed before we work with 1s of audio input sampled at 16kHz yielding 16,000 samples. 

\begin{figure}[h]
\centering
\includegraphics[width=8cm,height=3cm]{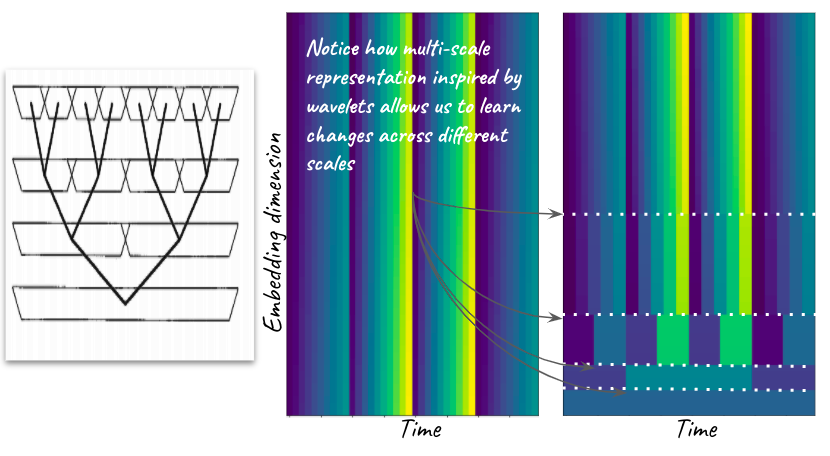}
\captionof{figure}{Core idea of wavelets utilizing multi-scale learning on (left) from \cite{berger1994removing}, and using them to create a layer that operates on intermediate Transformer embeddings at various scales. We show a demo signal and we retain half of them, and modify the other half using variable sized windows.
\label{fig:feature-graphic}}
\end{figure}

The window length is chosen to be 25ms, choosing a non-overlapping rectangular window. The rectangular window provides the network an optimal windowing function which, as we will see in a few of the learned filters, adapts itself to a shape of hanning/hamming window. We fix the front end to be a dense layer of size 2048 neurons followed by another layer of size 64, primarily to adapt to the size of the embedding layer of Transformer. A single dense layer of size 2048 successfully learned a filter-back to learn a neural-time frequency representation as shown in \cite{verma2016frequency}. This design was chosen as it produced state of the art results for an equally difficult problem of pitch estimation in polyphonic audio \cite{verma2016frequency}, with feed forward layers. Since Transformer layers consist only of attention + feed-forward blocks, we achieve an end-to-end architecture that does not have any convolutional operators.
This yields a front end representation of 40 time steps each of size 64 dimensions, (64 being a hyper-parameter). We choose 6 layers of Transformer module with the size of latent code being 64 for each of the layers, and 8 attention head, with 128 dim 3-layer dense layer to convert to the desired feature space. For comparing with a smaller model, we choose 3-layers of Transformers with similar setup. The last layer of Transformer is reduced to a smaller dimension using average pooling across time. The output of the last dense layer of dimension 200, chosen same as number of output labels.

\begin{figure*}[tb]
\centering
\includegraphics[width=13cm,height=4cm]{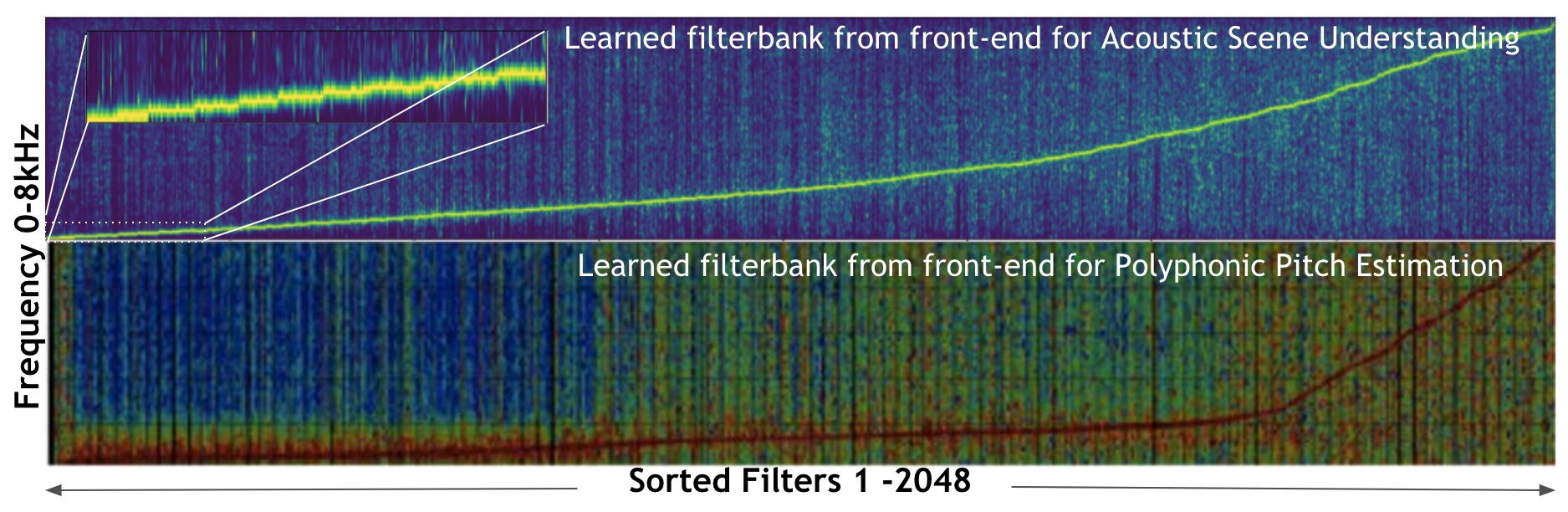}
\captionof{figure}{Sorted filters, learned by the front end, learns a problem specific non linear, non constant bandwidth filter-bank. This is shown by comparing it to  that learned by the same front end for polyphonic pitch estimation as shown in \cite{verma2016frequency}.
\label{fig:feature-graphic}}
\end{figure*}

\subsection{Transformer Architectures Inspired From CNNs: Pooling} We explored further performance enhancements to the baseline Transformer architecture proposed in the previous section. For this we draw inspiration of convolutional architectures used for the past decade to understand images \cite{deng2009imagenet} and audio \cite{gemmeke2017audio}. The traditional models e.g. Resnet-50 \cite{he2016deep}, consists of using a combination of convolutional layers followed by pooling. The use of pooling layers has two advantages. It reduces the number of computations by reducing the size of inputs in the higher layers. More importantly it allows the higher-layer neurons to have much broader receptive field sizes, and allows the network to learn hierarchically ordered features, which is important for a particular problem. Pooled Transformers outperform the baseline Transformer architecture, while retaining the same number of parameters. In our experiments average pooling performed significantly better than max-pooling, as it retains more information on the signal. As described in Figure 1, we use pooling across time after every two layers of Transformers, with stride 1, to reduce the dimensionality of the input by a factor of 2, and shows significant performance gain as compared to the original Transformer architecture without pooling. 

\subsection{Learning multi-scale embeddings}

In this adaptation, we draw inspiration from wavelet decomposition and success of pooling layers. We explored if we can decompose the intermediate embeddings out of the Transformer, at multiple scale similar to idea of wavelet decomposition. In order to achieve it, we fix up our kernel to be average operation across \textit{all} windows chosen at a particular level. Notice that we choose different widow sizes at different dimensions of embedding along the time axis.  The manner of implementation is again a design choice and there are several interesting ideas possible in future, including the choice of kernel. We draw inspiration from the work carried out in \cite{berger1994removing}, as seen in Figure 2. We adapt the window size, in factors of 1,2,4,8 and so on, following a geometric progression. The value is assigned to \textit{all} of the elements as opposed to reducing the size, as done in pooling thus retaining the same size. This operation in fully differentiable, and can be trained in end-to-end architectures. This is different than work carried out on spectral filtering \cite{tamkin2020language}, as we choose to operate firstly with variable window size as opposed to fixed windows, and secondly do not take explicit hand-crafted bands of filters. Additionally, we choose to model the space of embeddings-time hierarchically with only a few large windows, and large number of smaller window, most of them being 1 to retain the embeddings at their original scale. This retains the original transformer embeddings, with half of the embeddings unchanged, and tinkers with the other half. This combination has been at the core of wavelet transforms.

\section{Results \& Discussion}
For all of the architectures, we only tuned learning rate to be consistent with the results shown in \cite{fonseca2020fsd50k}. All of the Transformers have 6 layers (3 for small transformers) with 64 dim embeddings, and 3-Layer 128 neuron feed forward layers, and 8 attention head. The front end consists of 1024/2048 dimensional layer followed by a 64 dimensional dense layer for small and large transformers. We compared the same Transformer architectures with that of using i) pooling layers ii) multi-scale filters. We observed that even the smallest of the Transformer architectures outperform traditional convolutional architectures. This is quite significant,  unlike problems in vision \cite{dosovitskiy2020image}, where the margin was not as significant. Another observation is also that the performance keeps improving with more depth. All the models were trained using Tensorflow framework \cite{abadi2016tensorflow}, with Huber Loss as the error criteria between the predictions and the ground truth, using Adam optimizer \cite{kingma2014adam}. We see that the multi-scale approach outperforms the same transformer architecture without the intermediate layers, which is encouraging. However, is not able to beat the pooling layers. This may perhaps be due to ability of Transformers to better model smaller latent embeddings across time. 

\begin{table}[ht]
  \caption{\itshape Comparison of various proposed architecture as shown in the table below for mean average precision (mAP) metric. We see how even baseline Transformer architectures without using any convolutional layers can outperform widely used CNN architectures for acoustic scene understanding by significant margins. \cite{fonseca2020fsd50k} }
	\centering
	\begin{tabular}{|c|c|c|}
		\hline
		Neural Model Architecture & mAP & \# Param\\\hline
		CRNN \cite{fonseca2020fsd50k} & 0.417 & 0.96M\\
		VGG-like \cite{fonseca2020fsd50k} & 0.434 & 0.27M \\
		ResNet-18 \cite{fonseca2020fsd50k} & 0.373  & 11.3M\\
		DenseNet-121 \cite{fonseca2020fsd50k} & 0.425 & 12.5M \\\hline
		Small Transformer & 0.469 & 0.9M\\
		Large 6- Layer Transformer & \textbf{0.525}  & 2.3M \\
		Large Transformer with multi-scale filters & \textbf{0.541}  & 2.3M \\
		Large 6- Layer Transformer with Pooling &  \textbf{0.537} & 2.3M
        \\\hline
	\end{tabular}
	\label{tab:example}
\end{table}

\subsection{What the front end learns} We follow a strategy similar to that described in \cite{verma2016frequency} to understand what the filters learn. We deploy the same front end in our work which is again a feed-forward layer consisting of 2048 neuron, followed by a 64-dim dense layer. This is similar to the analogy of getting a mel-like representation which is learnable end-to-end. After training, we take the first layer and sort the filter according to the peaks of their Fourier representation. We see that it manages to learn a non-linear, non-constant bandwidth filterbank as seen in Figure 3. We also see that with using the same front end for two different applications, namely for pitch estimation and acoustic scene understanding, the shape and the resolution of the learned filter-bank is different. In addition, we can also see a step-wise pattern, which shows multiple filters assigned to the same frequency bin to account for the phase variations of the input signals. Figure 4 depicts a few chosen filters for the sake of discussion here. We observe a variety of ideas that can be interpreted from signal processing perspective, and also to take into account the characteristics of the input signal i.e. frequency, timbre, and energy. We can see,  in center-top row, that a filter learns a pure sinusoidal basis of a certain frequency. Furthermore, it also manages to learn a windowing function that closely resembles hanning/hamming window.  The filters in the left column present at the top-bottom are characteristic of an onset detector, which can, respectively, be a slow/rapid onset.  Further, the filter present in the second row, third column shows a slowly moving signal, which may be latching onto the overall energy envelop of a signal for certain characteristic sounds. These correlations to traditional signal processing ideas present in these filters are compelling. 

\begin{figure}[h]
\centering
\includegraphics[width=6.4cm,height=3cm]{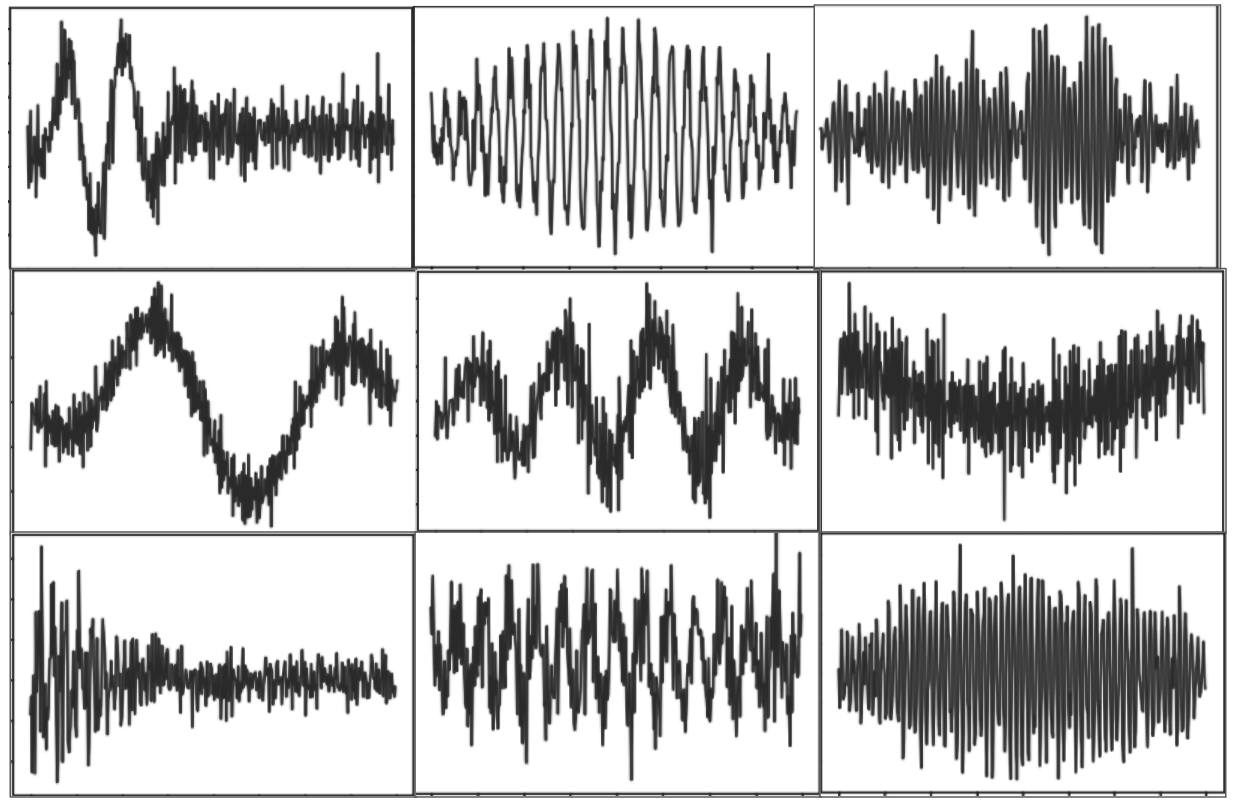}
\captionof{figure}{Filters learned from the first layer of front end show strong correlations to signal processing, particularly learning sinusoidal signals, onset detectors, energy envelops, and windowing functions}
\end{figure}
\section{Conclusion \& Future Work}
\label{sec:conclusion}
We have shown here how a Transformer architecture without the use of convolutional filters can be adapted for large scale audio understanding. The work is promising, outperforming other convolutional architectures by a significant margin. We show our model can learn a time frequency front end that is adaptable to the particular problem of interest, in this case, large scale audio understanding.  
There are several possible research directions ahead. With the advancements in Transformer architectures such as switch transformers \cite{fedus_zoph_shazeer_2021}, and sparse transformers \cite{child2019generating}, these results would further improve. Additionally, with the success of unsupervised representation learning architectures for audio \cite{verma2020framework}, it will be interesting to do large scale pre-training for making robust audio representations. It will be also be useful to explore a wider search over hyper-parameters to increase the reported precision scores, which most likely will go up even further.  

%\section{Acknowledgement}
%\label{sec:acknowledgement}
%The authors would like to express their gratitude to all the healthcare workers for working tirelessly to save countless lives during the pandemic. 

% -------------------------------------------------------------------------
% Either list references using the bibliography style file IEEEtran.bst
\bibliographystyle{IEEEtran}
\bibliography{refs21}

\begin{thebibliography}{10}
\providecommand{\url}[1]{#1}
\def\UrlFont{\rmfamily}
\providecommand{\newblock}{\relax}
\providecommand{\bibinfo}[2]{#2}
\providecommand\BIBentrySTDinterwordspacing{\spaceskip=0pt\relax}
\providecommand\BIBentryALTinterwordstretchfactor{4}
\providecommand\BIBentryALTinterwordspacing{\spaceskip=\fontdimen2\font plus
\BIBentryALTinterwordstretchfactor\fontdimen3\font minus
  \fontdimen4\font\relax}
\providecommand\BIBforeignlanguage[2]{{%
\expandafter\ifx\csname l@#1\endcsname\relax
\typeout{** WARNING: IEEEtran.bst: No hyphenation pattern has been}%
\typeout{** loaded for the language `#1'. Using the pattern for}%
\typeout{** the default language instead.}%
\else
\language=\csname l@#1\endcsname
\fi
#2}}

\bibitem{verma2020framework}
P.~Verma and J.~Smith, ``A framework for contrastive and generative learning of
  audio representations,'' \emph{arXiv preprint arXiv:2010.11459}, 2020.

\bibitem{he2016deep}
K.~He, X.~Zhang, S.~Ren, and J.~Sun, ``Deep residual learning for image
  recognition,'' in \emph{Proceedings of the IEEE conference on computer vision
  and pattern recognition}, 2016, pp. 770--778.

\bibitem{gemmeke2017audio}
J.~F. Gemmeke, D.~P. Ellis, D.~Freedman, A.~Jansen, W.~Lawrence, R.~C. Moore,
  M.~Plakal, and M.~Ritter, ``Audio set: An ontology and human-labeled dataset
  for audio events,'' in \emph{2017 IEEE International Conference on Acoustics,
  Speech and Signal Processing (ICASSP)}.\hskip 1em plus 0.5em minus
  0.4em\relax IEEE, 2017, pp. 776--780.

\bibitem{madani2020progen}
A.~Madani, B.~McCann, N.~Naik, N.~S. Keskar, N.~Anand, R.~R. Eguchi, P.-S.
  Huang, and R.~Socher, ``Progen: Language modeling for protein generation,''
  \emph{arXiv preprint arXiv:2004.03497}, 2020.

\bibitem{brown2020language}
T.~B. Brown, B.~Mann, N.~Ryder, M.~Subbiah, J.~Kaplan, P.~Dhariwal,
  A.~Neelakantan, P.~Shyam, G.~Sastry, A.~Askell, \emph{et~al.}, ``Language
  models are few-shot learners,'' \emph{arXiv preprint arXiv:2005.14165}, 2020.

\bibitem{devlin2018bert}
J.~Devlin, M.-W. Chang, K.~Lee, and K.~Toutanova, ``Bert: Pre-training of deep
  bidirectional transformers for language understanding,'' \emph{arXiv preprint
  arXiv:1810.04805}, 2018.

\bibitem{huang2018music}
C.-Z.~A. Huang, A.~Vaswani, J.~Uszkoreit, N.~Shazeer, I.~Simon, C.~Hawthorne,
  A.~M. Dai, M.~D. Hoffman, M.~Dinculescu, and D.~Eck, ``Music transformer,''
  \emph{arXiv preprint arXiv:1809.04281}, 2018.

\bibitem{sun2019videobert}
C.~Sun, A.~Myers, C.~Vondrick, K.~Murphy, and C.~Schmid, ``Videobert: A joint
  model for video and language representation learning,'' in \emph{Proceedings
  of the IEEE/CVF International Conference on Computer Vision}, 2019, pp.
  7464--7473.

\bibitem{girdhar2019video}
R.~Girdhar, J.~Carreira, C.~Doersch, and A.~Zisserman, ``Video action
  transformer network,'' in \emph{Proceedings of the IEEE/CVF Conference on
  Computer Vision and Pattern Recognition}, 2019, pp. 244--253.

\bibitem{dosovitskiy2020image}
A.~Dosovitskiy, L.~Beyer, A.~Kolesnikov, D.~Weissenborn, X.~Zhai,
  T.~Unterthiner, M.~Dehghani, M.~Minderer, G.~Heigold, S.~Gelly,
  \emph{et~al.}, ``An image is worth 16x16 words: Transformers for image
  recognition at scale,'' \emph{arXiv preprint arXiv:2010.11929}, 2020.

\bibitem{parmar2018image}
N.~Parmar, A.~Vaswani, J.~Uszkoreit, L.~Kaiser, N.~Shazeer, A.~Ku, and D.~Tran,
  ``Image transformer,'' in \emph{International Conference on Machine
  Learning}.\hskip 1em plus 0.5em minus 0.4em\relax PMLR, 2018, pp. 4055--4064.

\bibitem{dhariwal2020jukebox}
P.~Dhariwal, H.~Jun, C.~Payne, J.~W. Kim, A.~Radford, and I.~Sutskever,
  ``Jukebox: A generative model for music,'' \emph{arXiv preprint
  arXiv:2005.00341}, 2020.

\bibitem{verma2018neural}
P.~Verma and J.~O. Smith, ``Neural style transfer for audio spectograms,''
  \emph{arXiv preprint arXiv:1801.01589}, 2018.

\bibitem{verma2019neuralogram}
P.~Verma, C.~Chafe, and J.~Berger, ``Neuralogram: A deep neural network based
  representation for audio signals,'' \emph{arXiv preprint arXiv:1904.05073},
  2019.

\bibitem{oord2017neural}
A.~v.~d. Oord, O.~Vinyals, and K.~Kavukcuoglu, ``Neural discrete representation
  learning,'' \emph{arXiv preprint arXiv:1711.00937}, 2017.

\bibitem{baevski2020wav2vec}
A.~Baevski, H.~Zhou, A.~Mohamed, and M.~Auli, ``wav2vec 2.0: A framework for
  self-supervised learning of speech representations,'' \emph{arXiv preprint
  arXiv:2006.11477}, 2020.

\bibitem{baevski2019vq}
A.~Baevski, S.~Schneider, and M.~Auli, ``vq-wav2vec: Self-supervised learning
  of discrete speech representations,'' \emph{arXiv preprint arXiv:1910.05453},
  2019.

\bibitem{haque2018conditional}
A.~Haque, M.~Guo, and P.~Verma, ``Conditional end-to-end audio transforms,''
  \emph{arXiv preprint arXiv:1804.00047}, 2018.

\bibitem{haque2019audio}
A.~Haque, M.~Guo, P.~Verma, and L.~Fei-Fei, ``Audio-linguistic embeddings for
  spoken sentences,'' in \emph{ICASSP 2019-2019 IEEE International Conference
  on Acoustics, Speech and Signal Processing (ICASSP)}.\hskip 1em plus 0.5em
  minus 0.4em\relax IEEE, 2019, pp. 7355--7359.

\bibitem{vaswani2017attention}
A.~Vaswani, N.~Shazeer, N.~Parmar, J.~Uszkoreit, L.~Jones, A.~N. Gomez,
  L.~Kaiser, and I.~Polosukhin, ``Attention is all you need,'' \emph{arXiv
  preprint arXiv:1706.03762}, 2017.

\bibitem{fonseca2020fsd50k}
E.~Fonseca, X.~Favory, J.~Pons, F.~Font, and X.~Serra, ``Fsd50k: an open
  dataset of human-labeled sound events,'' \emph{arXiv preprint
  arXiv:2010.00475}, 2020.

\bibitem{virtanen2020scipy}
P.~Virtanen, R.~Gommers, T.~E. Oliphant, M.~Haberland, T.~Reddy, D.~Cournapeau,
  E.~Burovski, P.~Peterson, W.~Weckesser, J.~Bright, \emph{et~al.}, ``Scipy
  1.0: fundamental algorithms for scientific computing in python,''
  \emph{Nature methods}, vol.~17, no.~3, pp. 261--272, 2020.

\bibitem{opennmt}
\BIBentryALTinterwordspacing
G.~Klein, Y.~Kim, Y.~Deng, J.~Senellart, and A.~M. Rush, ``Opennmt: Open-source
  toolkit for neural machine translation,'' in \emph{Proc. ACL}, 2017.
  [Online]. Available: \url{https://doi.org/10.18653/v1/P17-4012}
\BIBentrySTDinterwordspacing

\bibitem{berger1994removing}
J.~Berger, R.~R. Coifman, and M.~J. Goldberg, ``Removing noise from music using
  local trigonometric bases and wavelet packets,'' \emph{Journal of the Audio
  Engineering Society}, vol.~42, no.~10, pp. 808--818, 1994.

\bibitem{verma2016frequency}
P.~Verma and R.~W. Schafer, ``Frequency estimation from waveforms using
  multi-layered neural networks.'' in \emph{INTERSPEECH}, 2016, pp. 2165--2169.

\bibitem{deng2009imagenet}
J.~Deng, W.~Dong, R.~Socher, L.-J. Li, K.~Li, and L.~Fei-Fei, ``Imagenet: A
  large-scale hierarchical image database,'' in \emph{2009 IEEE conference on
  computer vision and pattern recognition}.\hskip 1em plus 0.5em minus
  0.4em\relax Ieee, 2009, pp. 248--255.

\bibitem{tamkin2020language}
A.~Tamkin, D.~Jurafsky, and N.~Goodman, ``Language through a prism: A spectral
  approach for multiscale language representations,'' \emph{Advances in Neural
  Information Processing Systems}, vol.~33, 2020.

\bibitem{abadi2016tensorflow}
M.~Abadi, P.~Barham, J.~Chen, Z.~Chen, A.~Davis, J.~Dean, M.~Devin,
  S.~Ghemawat, G.~Irving, M.~Isard, \emph{et~al.}, ``Tensorflow: A system for
  large-scale machine learning,'' in \emph{12th $\{$USENIX$\}$ symposium on
  operating systems design and implementation ($\{$OSDI$\}$ 16)}, 2016, pp.
  265--283.

\bibitem{kingma2014adam}
D.~P. Kingma and J.~Ba, ``Adam: A method for stochastic optimization,''
  \emph{arXiv preprint arXiv:1412.6980}, 2014.

\bibitem{fedus_zoph_shazeer_2021}
\BIBentryALTinterwordspacing
W.~Fedus, B.~Zoph, and N.~Shazeer, ``Switch transformers: Scaling to trillion
  parameter models with simple and efficient sparsity,'' Jan 2021. [Online].
  Available: \url{https://arxiv.org/abs/2101.03961}
\BIBentrySTDinterwordspacing

\bibitem{child2019generating}
R.~Child, S.~Gray, A.~Radford, and I.~Sutskever, ``Generating long sequences
  with sparse transformers,'' \emph{arXiv preprint arXiv:1904.10509}, 2019.

\end{thebibliography}
%
% or list them by yourself
% \begin{thebibliography}{9}
% 
% \bibitem{waspaa21web}
%   \url{http://www.waspaa.com}.
%
% \bibitem{IEEEPDFSpec}
%   {PDF} specification for {IEEE} {X}plore$^{\textregistered}$,
%   \url{http://www.ieee.org/portal/cms_docs/pubs/confstandards/pdfs/IEEE-PDF-SpecV401.pdf}.
%
% \bibitem{PDFOpenSourceTools}
%   Creating high resolution {PDF} files for book production with 
%   open source tools, 
%   \url{http://www.grassbook.org/neteler/highres_pdf.html}.
%
% \bibitem{eWilliams1999}
% E. Williams, \emph{Fourier Acoustics: Sound Radiation and Nearfield Acoustic
%   Holography}. London, UK: Academic Press, 1999.
% 
% \bibitem{ieeecopyright}
%   \url{http://www.ieee.org/web/publications/rights/copyrightmain.html}.
%
% \bibitem{cJones2003}
% C. Jones, A. Smith, and E. Roberts, ``A sample paper in conference
%   proceedings,'' in \emph{Proc. IEEE ICASSP}, vol. II, 2003, pp. 803--806.
% 
% \bibitem{aSmith2000}
% A. Smith, C. Jones, and E. Roberts, ``A sample paper in journals,'' 
%   \emph{IEEE Trans. Signal Process.}, vol. 62, pp. 291--294, Jan. 2000.
% 
% \end{thebibliography}

\end{sloppy}
\end{document}